\documentclass[10pt]{npqcd}
\begin{document}
\renewcommand\nextpg{\pageref{pgs1}}\renewcommand\titleA{
 Just one more approach to calculating the QCD ground state
}\renewcommand\authorA{
 S. V. Molodtsov${}^{1,2,~\mathrm{a}}$,
 G. M. Zinovjev${}^{3,~\mathrm{b}}$
}\renewcommand\email{
 e-mail:\space \eml{a}{molodtsov@itep.ru},
\eml{b}{Gennady.Zinovjev@cern.ch}\\[1mm]
}\renewcommand\titleH{
 Just one more approach to calculating the QCD ground state
}\renewcommand\authorH{
 Molodtsov S.V., Zinovjev G.M.
}\renewcommand\titleC{\titleA}\renewcommand\authorC{\authorH}\renewcommand\institution{
$^1$Joint Institute for Nuclear Research, RU-141980, Dubna, Moscow region, RUSSIA.\\
$^2$Institute of Theoretical and Experimental Physics, RU-117259, Moscow, RUSSIA.\\
$^3$Bogolyubov Institute for Theoretical Physics, UA-03143, Kiev, UKRAINE.
}\renewcommand\abstractE{
 The quark behaviour in the background of intensive stochastic gluon field is
studied. An approximate procedure for calculating the effective Hamiltonian is
developed and the corresponding ground state within the Hartree-Fock-Bogolyubov
approach is found. The comparative analysis of various model Hamiltonian is given
and transition to the chiral limit in the Keldysh model is discused in detail.
We study response to the process of filling up
the Fermi sphere with quarks, calculate the vacuum pressure and demonstrate
the existence of filled-in state degenerate with the vacuum one.
}
\begin{article}

We study the quark (anti-quark) behaviour while being influenced by intensive stochastic
gluon field and work in the context of the Euclidean field theory. The corresponding
Lagrangian density is the following
\begin{equation}
\label{1}
{\cal L}_E=\bar q~(i\gamma_\mu D_\mu+im)~q~,
\end{equation}
here  $q$ ($\bar q$) --- are the quark (anti-quarks) fields with covariant derivative
$D_\mu=\partial_\mu -i g A^a_\mu t^a$ where $A^a_\mu$ is the  gluon field, $t^a=\lambda^a/2$
are the generators of colour gauge group $SU(N_c)$ and $m$ is the current quark mass.
As the model of stochastic gluon field we refer to the example of (anti-)instantons considering
an ensemble of these quasi-classical configurations. On the way to construct an effective theory
we consider the quenched approximation and neglect all the contributions coming from gluon fields
$A_{ex}$ generated by the (anti-)quarks $A_{ex} \ll A$.
Then the corresponding Hamiltonian description results from
\begin{equation}
\label{2}
{\cal H}=\pi \dot q-{\cal L}_E~,~~\pi=\frac{\partial{\cal L}_E}{\partial \dot q}=i q^{+}~,
\end{equation}
and ${\cal H}_0=-\bar q~(i{\bf \gamma}{\bf \nabla}+im)q$,
for noninteracting quarks.
In Schr\"odinger representation the quark field evolution is determined by the equation for
the quark probability amplitude $\Psi$ as
\begin{equation}
\label{4}
\dot \Psi=- H \Psi~,
\end{equation}
with the density of interaction Hamiltonian
${\cal V}_S=\bar q({\bf x})~t^a\gamma_\mu A^{a}_\mu(t,{\bf x})~q({\bf x})$.
The explicit dependence on "time" is present at the gluon field only. The creation and annihilation
operators of quarks and anti-quarks $a^+, a$, $b^+, b$ have no "time" dependence and consequently
\begin{equation}
\label{5}
q_{\alpha i}({\bf x})=\int\frac{d {\bf p}}{(2\pi)^3} \frac{1}{(2|p_4|)^{1/2}}~
\left[~a({\bf p},s,c)~u_{\alpha i}({\bf p},s,c)~ e^{i{\bf p}{\bf x}}+
b^+({\bf p},s,c)~v_{\alpha i}({\bf p},s,c)~ e^{-i{\bf p}{\bf x}}\right]~.
\end{equation}
The stochastic character of gluon field (which we supposed) allows us to develop the approximate
description of the state $\Psi$ if the following procedure of averaging
$\Psi \to \langle \Psi \rangle=\int_0^t ~d\tau ~\Psi(\tau)/~t$
is intoduced. With this procedure taken the futher step is to turn to the approach of constructing
a density matrix $\langle \stackrel{*}{\Psi}\stackrel{}{\Psi} \rangle$. However, here we believe
that at calculating the ground state (or more generally with quasi-stationary state) it might be
sufficiently informative to operate with the averaged amplitude directly. Then in the interaction
representation  $\Psi=e^{H_0 t}\Phi$ we have the equation for state $\Phi$ as
$\dot \Phi=- V \Phi$, $V=e^{H_0 t} V_S e^{-H_0 t}$.
Now the "time" dependence appears in quark operators as well and after averaging over the
short-wavelength component one may obtain the following equation
\begin{equation}
\label{8}
\langle\dot\Phi(t)\rangle=+\int_0^{\infty} d\tau ~\langle V(t) V(t-\tau)\rangle~\langle\Phi(t)\rangle~.
\end{equation}
The limitations to have such a factorization validated are well known in the theory of stochastic
differential equations (see, for example, \cite{van}). The integration interval in Eq.(\ref{8}) may be
extended to the infinite "time" because of the (supposed) rapid decrease of the corresponding
correlation function. Now we are allowed to deal with amplitude $\langle\Phi(t)\rangle$ in the right
hand side of Eq.(\ref{8}) instead the amplitude with the shifted arguments in order to get an ordinary
integro-differential equation. In the quantum field theory applications it is usually difficult to
construct the correlation function in the most general form. However, if we are going to limit our
interest by describing the long-wavelength quark component only then gluon field correlator
$\langle A^a_\mu(x) A^b_\nu(y)\rangle$ may be factorized and as a result we have
$$
\langle\dot\Phi(t)\rangle=\int d{\bf x}~ \bar q({\bf x},t)~t^a\gamma_\mu~q({\bf x},t)~
\int_0^{\infty} d\tau \int d{\bf y} ~\bar q({\bf y},t-\tau)~t^b\gamma_\nu~q({\bf y},t-\tau)~
g^2\langle A^{a}_\mu(t,{\bf x}) A^{b}_\nu(t-\tau,{\bf y})\rangle~
\langle\Phi(t)\rangle~.$$
Having assumed the correlation function rapidly decreasing in "time" we could ignore all the
retarding effects in the quark operators.
Turning back to the Schr\"odinger representation we have for the state amplitude
$\chi=e^{-H_0 t}\langle\Phi\rangle$ the following equation
\begin{equation}
\label{9}
\dot \chi=- H_{ind}~ \chi~,
~~~{\cal H}_{ind}=-\bar q~(i{\bf \gamma}{\bf \nabla}+im)~q-\bar q~t^a\gamma_\mu~q~
  \int d{\bf y} ~\bar q'~t^b\gamma_\nu~q'~
\int_0^{\infty} d\tau~ g^2\langle A^{a}_\mu A^{'b}_\nu\rangle~,
\end{equation}
with $q=q({\bf x})$, $\bar q=\bar q({\bf x})$, $q'=q({\bf y})$, $\bar q'=\bar q({\bf y})$ and
$A^{a}_\mu =A^{a}_\mu(t,{\bf x})$, $A^{'b}_\nu=A^{b}_\nu(t-\tau,{\bf y})$. Now the correlation
function might be presented as
$\int_0^{\infty} d\tau~g^2\langle A^{a}_\mu A^{'b}_\nu\rangle=
\delta^{a b}~\delta_{\mu\nu}~I({\bf x}-{\bf y})+J_{\mu\nu}({\bf x}-{\bf y})$.
In our consideration we ignore the contribution of the second formfactor spanning on the components
of the vector ${\bf x}-{\bf y}$. Thus, on output we receive the Hamiltonian
of four-fermion interaction with the formfactor rooted in the presence of two quark currents
in the points ${\bf x}$ and ${\bf y}$.
With this form of the effective Hamiltonian we could apply the Hartree--Fock--Bogolyubov
method  to find its ground state as one constructed by the quark--anti-quark pairs
with the oppositely directed momenta
\begin{equation}
\label{10}
|\sigma\rangle=T~|0\rangle~,~~~
T=\Pi_{ p,s,c}~\exp\left\{~\frac{\theta}{2}~\left[~a^+({\bf p},s,c)~b^+(-{\bf p},s,c)+
a({\bf p},s,c)~b(-{\bf p},s,c)~\right]~\right\}~,
\end{equation}
where the parameter $\theta({\bf p})$ characterizes the pairing strength.
Minimizing the mean energy
functional one is able to determine the angle $\theta$ magnitude
$\frac{d\langle\sigma|H_{ind}|\sigma\rangle}{d\theta}=0$.
By the help of dressing tranformation $T$ we introduce the creation
and annihilation operators of quasi-particles
$A=T~a~T^{-1}$, $B=T~b~T^{-1}$.
Dropping the calculation details out (see Ref. \cite{MZ})
we present here the following result for the mean energy
as a function of the $\theta$ angle
\begin{eqnarray}
\label{26}
&&\langle\sigma|H_{ind}|\sigma\rangle=-\int \frac{d {\bf p}}{(2\pi)^3}~\frac{2N_c~p_4^{2}}{|p_4|}
\left(1-\cos\theta\right)-
\nonumber\\
&&-\widetilde G\int \frac{d {\bf p}d {\bf q}}{(2\pi)^6}\left\{-(3\widetilde I-\widetilde J)
\frac{p_4~ q_4}{|p_4||q_4|}+(4\widetilde I-\widetilde J)\frac{p~ q}{|p_4||q_4|}
\left(\sin\theta-\frac{m}{p}\cos\theta\right)
\left(\sin\theta'-\frac{m}{q}\cos\theta'\right)+\right.\nonumber\\ [-.2cm]
\\ [-.25cm]
&&
\left.+(-2\widetilde I\delta_{ij}-2\widetilde J_{ij}+\widetilde J\delta_{ij})~\frac{p_i~q_j}{|p_4||q_4|}~
\left(\cos\theta+\frac{m}{p}\sin\theta\right)
\left(\cos\theta'+\frac{m}{ q}\sin\theta'\right)~\right\}~,\nonumber
\end{eqnarray}
here the following designations are used $p=|{\bf p}|$, $q=|{\bf q}|$,
$\widetilde I=\widetilde I({\bf p}+{\bf q})$, $\widetilde J_{ij}=\widetilde J_{ij}({\bf p}+{\bf q})$,
$\widetilde J=\sum_{i=1}^3\widetilde J_{ii}$,
$p^2=q^2=-m^2$, $\theta'=\theta(q)$ where $\widetilde G$ is the constant of corresponding
four-fermion interaction (the relevant details can be found in \cite{MZ}).
The first integral in Eq. (\ref{26}) comes from free Hamiltonian, and we make
a natural subtraction (adding the unit)
in order to have zero mean free energy when the angle of pairing is trivial.
\begin{figure}
\begin{minipage}{.45\textwidth}
\centering
\includegraphics[bb= 0 0 600 600, width=.9\textwidth]{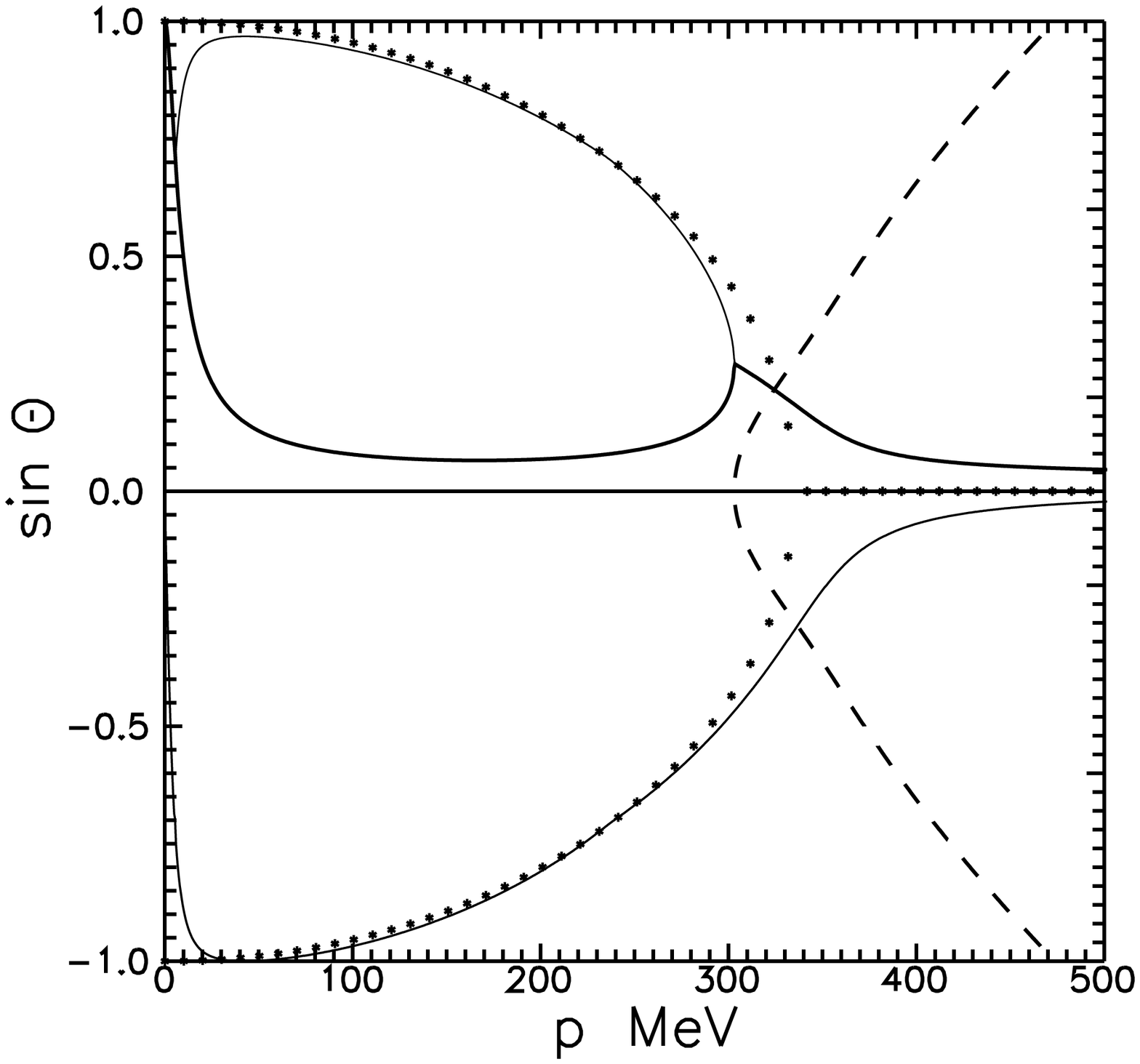}
\caption{Phase portrait of the Keldysh model, $\sin\theta$ as a function of momentum $p$(MeV)
(dashed line for imaginary values). The dotted
curves  corresponds to the solution  in the chiral limit $m=0$.}
\label{f1}
\end{minipage}
 \rule{.05\textwidth}{0pt}
\begin{minipage}{.45\textwidth}
\centering
\includegraphics[bb= 0 0 600 600 ,width=.9\textwidth]{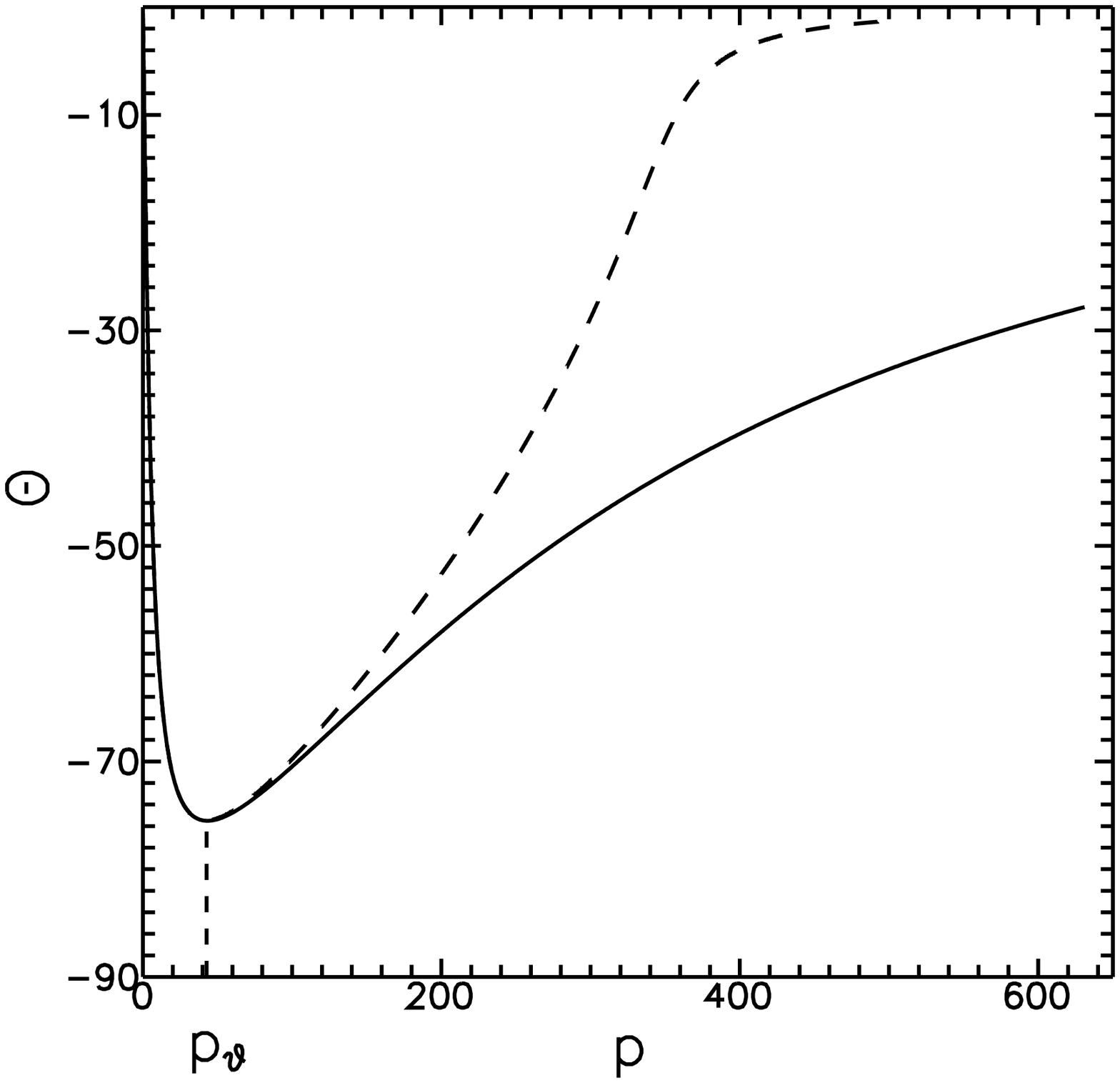}
\caption{The optimal angle $\theta$ as a function of momentum $p$(MeV). The solid line corresponds
to the NJL model and the dashed one to the Keldysh model. The current quark mass is $m=5.5$ MeV and
$p_\theta\sim 40$ MeV.}
\label{f2}
\end{minipage}
\end{figure}
{\bf Nambu--Jona-Lasinio model}.
In order to get an idea of the parameter scales we continue with handling the model in which the formfactor
behaves in the coordinate space as
$I({\bf x}-{\bf y})=\delta({\bf x}-{\bf y})$, $J_{\mu\nu}=0$, dropping contribution
spanned on the $p_i q_j$ tensor also. Actually, it corresponds to
the Nambu--Jona-Lasinio model \cite{njl}. As well known the model with such a formfactor
requires the regularization and, hence, the cutoff parameter $\Lambda$ comes to the play
\begin{equation}
\label{28}
W=\int^{\Lambda}\frac{d{\bf p}}{(2\pi)^3}~\left[|p_4|\left(1-\cos\theta\right)-
G\frac{p}{|p_4|}\left(\sin\theta-\frac{m}{p}\cos\theta\right)\int^\Lambda \frac{d{\bf q}}{(2\pi)^3}
\frac{q}{|q_4|}\left(\sin\theta'-\frac{m}{q}\cos\theta'\right)\right]~.
\end{equation}
We adjust the NJL model with the parameter set given by Hatsuda and Kunihiro \cite{njl} in which
$\Lambda=631 \mbox{MeV}$, $m=5.5\mbox{MeV}$. One curious point of this model is that the solution
for optimal angle $\theta$ in the whole interval $p\in[0,\Lambda]$ can be found by solving
the simple trigonometrical equation
$(p^2+m^2)~\sin\theta-M_q\left(p\cos\theta+m\sin\theta\right)=0$,
with the dynamical quark mass
$M_q=2G~\int^\Lambda \frac{d{\bf p}}{(2\pi)^3}\frac{p}{|p_4|}
~\left(\sin\theta-\frac{m}{p}\cos\theta\right)$.
Eventually the results obtained look like $M_q=-335$ MeV for dynamical quark mass and
$\langle\sigma|\bar q q|\sigma\rangle=-i~(245$ MeV$)^3$ for the quark condensate with the following
definition of the quark condensate
$\langle\sigma|\bar q q|\sigma\rangle=
\frac{i~N_c}{\pi^2}~\int_0^\infty dp~\frac{p^2}{|p_4|}~(p\sin\theta-m\cos\theta)$.
{\bf The Keldysh model}.
Now we are going to analyse the limit, in some extent, opposite to the NJL model, i.e. we are
dealing with the formfactor behaving as a delta function but in the momentum space
(analogously the Keldysh model, well known in the physics of condensed matter \cite{K}),
$I({\bf p})=(2\pi)^3~\delta({\bf p})$. Here the mean energy functional has the following form
\begin{equation}
\label{31}
W(m)=\int \frac{d{\bf p}}{(2\pi)^3}~\left[|p_4|~\left(1-\cos\theta\right)-
G~\frac{p^2}{|p_4|^2}\left(\sin\theta-\frac{m}{p}\cos\theta\right)^2\right]~.
\end{equation}
contrary to the NJL model there is no need to introduce any cut off.
The equation for calculating the optimal angle $\theta$ becomes the transcendental one
$|p_4|^3~\sin\theta-2G~\left(p\cos\theta+m\sin\theta\right)
\left(p\sin\theta-m\cos\theta\right)=0$,
and, clearly, it is rather difficult to get its solution in a general form. Fortunately, it is
much easier and quite informative to analyse the model in the chiral limit $m=0$.
There  exist one trivial solution $\theta=0$ and two nontrivial ones (for the positive and
negative angles) which obey the equation
$\cos\theta=\frac{p}{2G}$.
Obviously, these solutions are reasonable if the momentum is limited by $p<2G$. Then for
the mean energy (for real solution) we have:
$W_\pm(0)=-\frac{G^4}{15\pi^2}$, and for the quark condensate:
$\langle\sigma|\bar q q|\sigma\rangle(0)=\frac{i~N_c~G^3}{2\pi}$.
For the trivial solution the mean energy equals to zero together with the quark condensate
$W_0(0)=0$, $\langle\sigma|\bar q q|\sigma\rangle_0(0)=0$. Introducing the practical designation
$\sin\theta=\frac{M_\theta}{(p^2+M^2_\theta)^{1/2}}$ which characterizes the pairing strength
by the parameter $M_\theta$ we have, for example, for the nontrivial solution
$M_\theta=\left(4G^2-p^2\right)^{1/2}$. In order to compare the results with the NJL model
we fixed the value of four-fermion interaction constant as $M_\theta(0)=2G=335$ MeV.
It is interesting to notice that the respective energy becomes constant
$E(p)=\sqrt{p^2+M_\theta^2}$, $E(p)=2G$.

After having done the analysis in the chiral limit which is shown by the dotted line in Fig.1 we
would like to comment the situation beyond this limit. The evolution of corresponding branches is
available on the same plot \ref{f1}.  The
minimum of mean energy functional can be realized with the piecewise continuous functions.
At the local vicinity of coordinate origin we start with some branch of the solution,
then relevant solution passes from one possible branch to another one at any subinterval. But in any case
there is only one way to continue the real solution at streaming to the infinite limit.
As to the functional (\ref{31}) the
contribution of the term proportional to the cosine in the second parenthesis is divergent even if the
angle $\theta$ is zero. It means the mean energy out of chiral limit goes to an infinity at any
nonzero value of quark mass. The same conclusion is valid for
the chiral condensate. In principle this functional could be regularized and corresponding continuation
might be done but it is out of this presentation scope (see  Ref. \cite{MZ}).
It is not difficult to demonstrate the
similar discontinuities of functional are present, for example, for Gaussian
$
I({\bf x})=G~\exp{(-a^2~{\bf x}^2)}$,
and exponential
$
I({\bf x})=G~\exp{(-a~|{\bf x}|)}$,
formfactors and they are present even in the NJL model but this fact is masked by the cut off parameter.
Comparing the optimal angles in the NJL and Keldysh models (see Fig. 2) it is interesting to notice that the
formation of quasiparticles becomes significant at some momentum value close to the origin
$p_\theta\sim 40$ MeV (for the Gaussian and exponentional formfactors it is around $p_\theta\sim 150$ MeV)
but not directly at the zero value. It is clear the inverse value of this parameter
determines the characteristic size of quasiparticle.
Analysing the discontinuity of mean energy functional and quark condensate we face some
troubles at fitting the quark condensate, for example. However, the dynamical quark mass and quark
condensate are nonobservable quantities and it is curious to remark here that although the mean
energy of the quark system is minus infinity the meson observables are finite and even
in Keldysh model the mesons are recognizable with reasonable scale
and we can in principle make a fit for this observables \cite{mvz}.

\begin{figure}
\begin{minipage}{.45\textwidth}
\centering
\includegraphics[bb= 0 0 600 600, width=.9\textwidth]{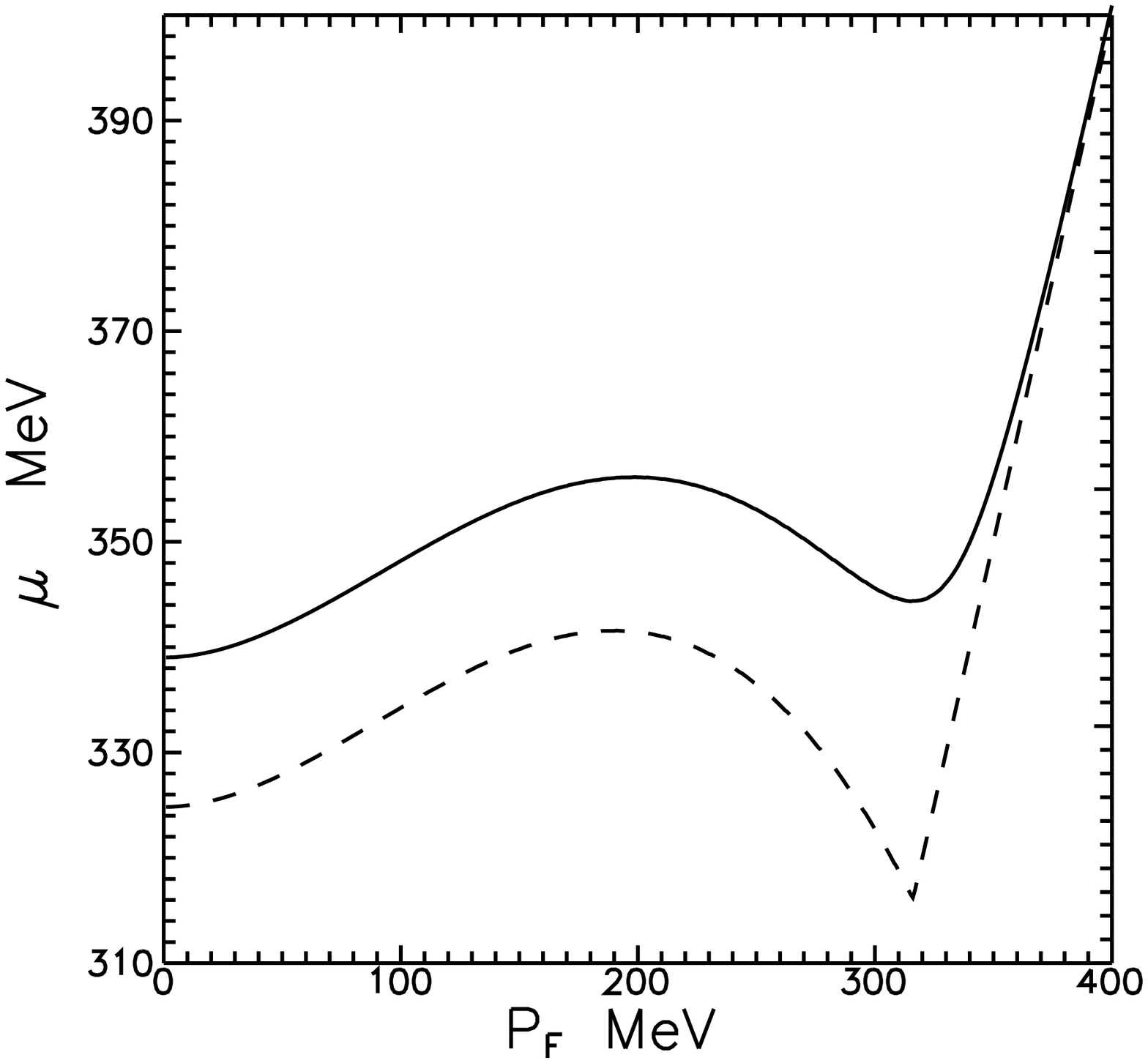}
\caption{The quark chemical potential as a function of the Fermi momentum for
the NJL model. The solid line corresponds to the current quark mass $m=5.5$ MeV
and the dashed one shows the behaviour in the chiral limit.}
\label{f3}
\end{minipage}
 \rule{.05\textwidth}{0pt}
\begin{minipage}{.45\textwidth}
\centering
\includegraphics[bb= 0 0 600 600 ,width=.9\textwidth]{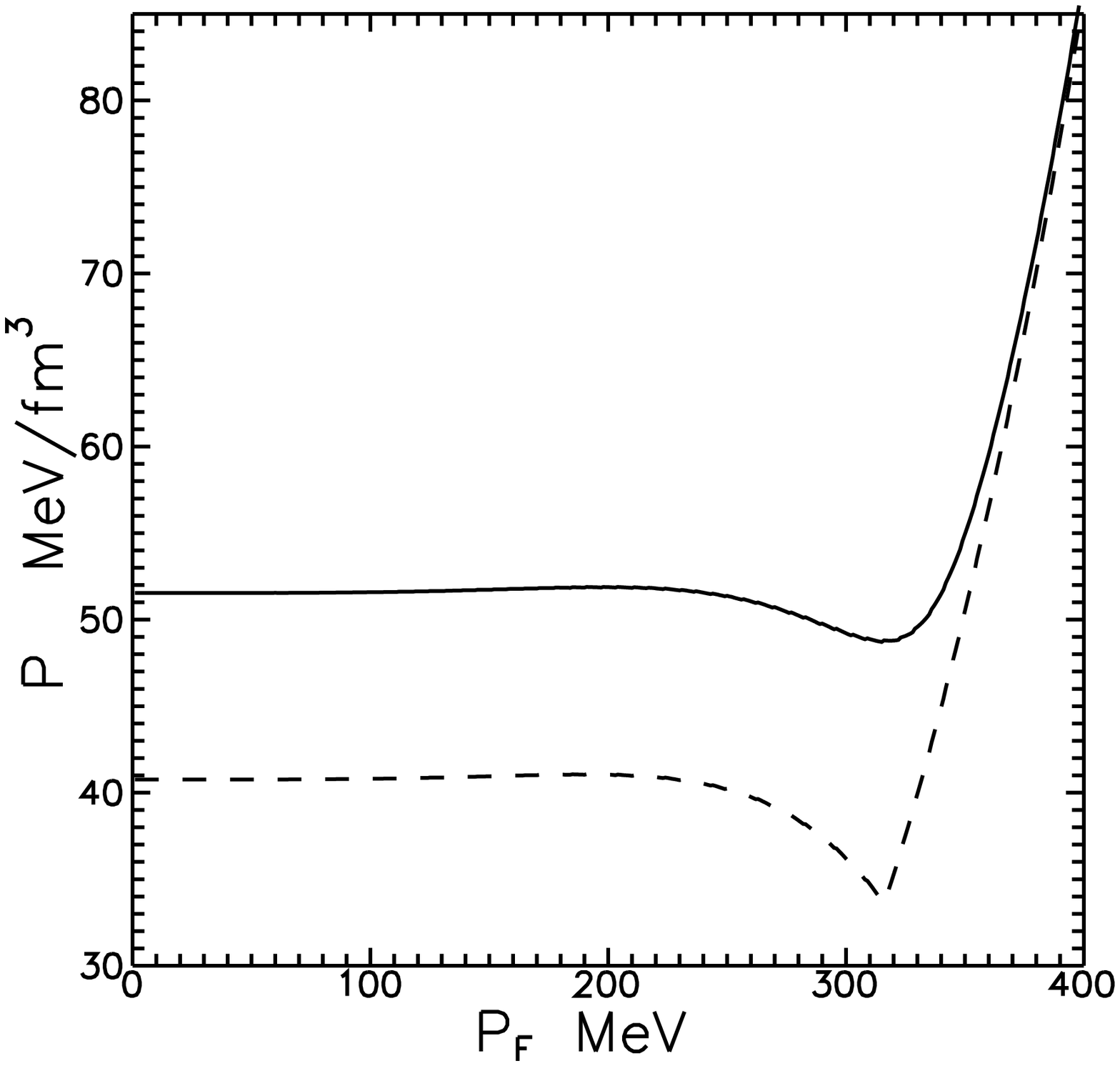}
\caption{The pressure of the quark ensemble as function of the Fermi momentum.
The solid line for the current quark mass $m=5.5$ MeV. The dashed one---in
chiral limit.}
\label{f4}
\end{minipage}
\end{figure}
Now our central issue could be formulated in the following way --- to
construct the state filled in by quasi-particles (the Sletter determinant)
$|N\rangle=\prod_{|{\mbox{\scriptsize{\bf p}}}|<P_F;s}~
A^+({\bf P};s)~|\sigma\rangle$,
which possesses the minimal mean energy $\langle N|H|N\rangle$ (surely, we
assume the quasi-particles are stable).
Here $P_F$ stands for the Fermi momentum and the polarization
runs over all possible values. It allows us to optimize the dressing
transformation and, as the consequence, to follow up the modifications of quasiparticles
being influenced by the process of filling in the Fermi sphere.
Eventualy it fixes the form of charge operator (particle number operator)
$|\langle N|\bar q i\gamma_4 q| N\rangle|$.
Let us define the partial energy density per one quark degree of freedom, as
$w=\frac{{\cal E}}{2 N_c}$, ${\cal E}=E/V$
where $E$ is the total energy of ensemble.
For the ensemble of quasi-particles
we obtain the following expression for the partial energy
\begin{eqnarray}
\label{27}
&&\langle N|w| N\rangle=\int^{P_F}\!\!\!\!\frac{d {\bf p}}{(2\pi)^3} |p_4| +
\int_{P_F}\frac{d {\bf p}}{(2\pi)^3} |p_4|(1-\cos\theta)-\nonumber\\[-.2cm]
\\ [-.25cm]
&&~~~~~~~~~~~~-
G\int_{P_F} \frac{d {\bf p}}{(2\pi)^3}~\frac{p}{|p_4|}~
\left(\sin\theta-\frac{m}{p}\cos\theta\right)
\int_{P_F} \frac{d {\bf q}}{(2\pi)^3}~\frac{q}{|q_4|}~ \left(\sin\theta'-
\frac{m}{q}\cos\theta'\right)~\widetilde I~.\nonumber
\end{eqnarray}
It could have ruther interesting interpretation if compared to the vacuum mean
energy Eq.(\ref{26}).
It is easy to see that for the state with the filled-in Fermi sphere the angles
of pairing could be defined by the condition of functional minimum (\ref{27}) only for the momenta
larger than Fermi momentum $P_F$. Then the quarks composing the Fermi sphere look like the
free (non-interacting) ones, as seen from the first term of Eq. (\ref{27}).
Now let us calculate the quark chemical potential which, by definition, is an
energy necessary for adding (removing) one quasi-particle to (from) a system
$\mu=\frac{\partial E}{\partial N}$, where
$N=2 N_c~ V~\int^{P_F}\frac{d {\bf p}}{(2\pi)^3}=\frac{N_c}{3\pi^2}~V~P_F^3$
is the total number of particles in the volume $V$.
Redefining the chemical potential as
$\mu=\frac{2\pi^2}{P_F^2}~\frac{\partial w}{\partial P_F}$
we consider the model with correlation function behaving as the
$\delta$-function in the coordinate space, which corresponds to NJL model.
The following relation could be obtained in this case (see Ref. \cite{fs})
$$\mu=[P_F^2+M_q^2]^{1/2}~.$$
Let us remind that for the free fermion gas the chemical potential increases
monotonically with the Fermi momentum growing. The curious feature of the NJL
model is the appearance of state almost degenerate with the vacuum state while
the process of filling up the Fermi sphere reaches to the momenta close to the
dynamical quark mass value (the similar value is peculiar to
the momentum of quark inside a baryon), see Fig. {\ref{f3}}.
This state density with the factor $3$ (which expresses the relation between
baryonic and quark degrees of freedom) absorbed corresponds to a normal nuclear
density ($n\sim 0.12$/fm$^3$), and chiral condensate could be estimated as
$|<\bar q q>^{1/3}|\sim 100$ MeV. In the chiral limit the chemical potential
is close to the discussed point and is even smaller than the vacuum one.
The full coincidence of the chemical potentials occurs at
the values of current quark mass around $2$ MeV. In fact, Fig. {\ref{f3}} shows that the
$u$ quark bond looks stronger than one of the $d$ quark.
The pressure of the quark ensemble
$P=-\frac{d E}{d V}=-\frac{\partial E}{\partial V}+\frac{P_F}{3V}~
\frac{\partial E}{\partial P_F}= -{\cal E}+\mu~n$,
is depicted in Fig. \ref{f4} as a function of the Fermi momentum where
$n=N/V$ is the quark density. The quark pressure at the values of the Fermi
momentum close to the quantity of dynamical quark mass is approximately
degenerate with the vacuum pressure (slightly lower than the vacuum one).
The vacuum density is of order  $40$---$50$ MeV/fm$^3$ and
corresponds well to the value extracted from the bag models.
Apparently our estimate of the effects responding to the process
of filling up the Fermi sphere entails a hope to understand a
routine feature of hadron world, namely, the fact of quark
equilibrium in the vacuum and inside the proton. The chemical
potential degeneracy and specific behaviour of the quark pressure
(with one new essential element which is just the presence
of instability region $dP/dP_F<0$) justify, in principle, the
conventional bag model. It urges to consider the filled states $|N\rangle$
as natural 'building' material for baryon octet (on the strong
interaction scale only).

\acknowledge{Acknowledgements}{We are grateful to the Organizers and personally
professor V. Skalozub for a well organized meeting.
This work was supported by the INTAS Grant 04-84-398 and the Grant
of National Academy of Sciences of Ukraine.
}

\end{article}
\label{pgs1}
\end{document}